\documentclass[12pt,preprint]{aastex}
\usepackage{emulateapj5,epsf}

\usepackage{lscape}

\shorttitle{
Exploring the disk-jet connection in YRL AGNs
}
\shortauthors{Kawakatu, Nagao and Woo}

\begin{document}

\title{
Exploring the disk-jet connection from the properties of 
narrow line regions in powerful young radio-loud AGNs
}


\author{Nozomu Kawakatu\altaffilmark{1}}
\affil{National Astronomical Observatory of Japan, 2-21-1 Osawa, 
Mitaka, Tokyo 181-8588, Japan}

\author{Tohru Nagao}
\affil{ Department of Physics, Graduate School of Science and Engineering, Ehime University, 2-5 Bunkyo-cho, Matsuyama 790-8577, Japan.
}

\author{Jong-Hak Woo\altaffilmark{2}}
\affil{Physics and Astronomy Department, University of California Los Angeles, 
CA 90095-1547}


\altaffiltext{1}{kawakatu@th.nao.ac.jp}
\altaffiltext{2}{Hubble Fellow}

\begin{abstract}
We investigate the optical emission-line flux ratios of narrow-line 
regions, in order to determine whether the formation of AGN jets 
requires specific accretion conditions.
We find that bright compact radio galaxies, which are powerful radio galaxies 
in the early stage of the jet activity, 
exhibit systematically larger flux ratios of [O{\sc i}]$\lambda$6300/[O{\sc iii}]$\lambda$5007 and smaller flux ratios of [O{\sc iii}]$\lambda$5007/[O{\sc iii}]$\lambda$4363 than radio-quiet (RQ) Seyfert 2 galaxies. 
Comparing the observed line ratios with photoionization models, 
it is found that the difference in the flux ratio of low- to high-ionization 
lines (e.g., [O{\sc i}]$\lambda$6300/[O{\sc iii}]$\lambda$5007) 
can be well understood by the difference in the spectral energy distribution 
(SED) of ionizing sources.
Powerful young radio-loud (YRL) AGNs favor SED without a strong big blue bump, 
i.e., a radiatively inefficient accretion flow (RIAF), while RQ AGNs are 
consistent with the models adopting SED with a strong big blue bump, i.e., 
a geometrically thin, optically thick disk. 
These findings imply that the formation of powerful AGN jets 
requires the accretion disk with harder ionizing SED (i.e., a RIAF). 
We discuss the obscuring structure of YRL AGNs as a plausible origin of 
the difference in flux ratios of [O{\sc iii}]$ \lambda$5007/[O{\sc iii}]$\lambda$4363. 

\end{abstract}

\section{Introduction} 
The formation of relativistic jets is one of the fundamental issues in
active galactic nuclei (AGNs) physics. The question how the jets form is
relevant to supermassive black hole (BH) cases as well as X-ray binaries 
(XRBs; see reviews by Fender 2003). Thanks to their short dynamical timescale, 
XRBs in various states have been observed in detail, and it is found that 
individual sources occupy particular accretion state with various X-ray luminosity
(e.g., Fender et al. 2004; Remilland \& McClintock 2006). 
These observations suggested that at the low luminosity (i.e., the 
low mass accretion rate) 
the jet power is a monotonic function of the accretion luminosity. 
On the other hand, the jet power may change abruptly by a large factor, 
following transitions between the soft state (a spectra characterized 
by the black body components) and hard state (a spectra characterized 
by the power-law components) at the high luminosity (i.e., the high 
mass accretion rate). 
By analogy to XRBs, it is expected that the power of AGN jets 
also depends on the mass accretion rates in a similar way. 
For example, Rawlings \& Saunders (1991) found that for the extended 
radio galaxies whose projected linear size ($LS$) is larger than $10$ kpc 
the jet power is proportional to the optical narrow line 
luminosity of NLRs which are photoionized by the nuclear continuum radiation 
(see also Willott et al. 1999). 
Ghisellini \& Celotti (2001) claimed that the less powerful radio galaxies such as Fanaroff-Riley type I radio galaxies (FR Is) exhibit the low accretion rate, while the powerful radio 
galaxies such as Fanaroff-Riley type I\hspace{-.1em}I radio galaxies (FR I\hspace{-.1em}Is) show the high accretion rate. 
This work is consistent with a scenario (e.g., Marchesini, Celotti, 
\& Ferrarese 2004; Wold et al. 2007; Hardcastle et al. 2007)
that FR Is have a radiatively inefficient accretion disk (RIAF) (e.g., Narayan \& Yi 1995), while FR I\hspace{-.1em}Is have a standard accretion disk 
(e.g., Shakura \& Sunyaev 1973).

In investigating the disk-jet connection in AGNs, however, we should keep 
in mind the relatively longer life time of the extended radio sources, i.e., FR Is and FR I\hspace{-.1em}Is. The reasons are as follows. 
The time-scale of the transition to the different physical states of 
the accretion disk is $\Delta t({\rm XRBs})\approx$ $10^{-2}-10^{-1}\, 
{\rm yr}$ (a few days $-$ a few weeks change) for XRBs 
(e.g., Fender et al. 2004). 
Assuming $M_{\rm BH}({\rm XRBs})=10M_{\odot}$ and $M_{\rm BH}({\rm AGNs})=
10^{8}M_{\odot}$, the transition time-scale for AGNs is 
$\Delta t({\rm AGNs})\approx 10^{5-6}\, {\rm yr}$
because of $\Delta t \propto M_{\rm BH}$. 
Thus, the physical states of the accretion disk may change in less than 
the typical age of extended radio galaxies, i.e., $t_{\rm age}\approx 10^{7-8}$ yr (e.g., Parma et al. 1999; Bird, Martini, \& Kaiser 2008). 
In other words, the observed relation between the jet power and the 
accretion power for the extended radio galaxies
may not be proving the direct disk-jet connection in AGNs. 
Thus, in order to investigate the disk-jet connection in AGNs more directly, 
it is essential to examine the physical states of the accretion disk in younger radio galaxies. 

The young and compact radio galaxies such as compact 
symmetric objects (CSOs; $LS <$ 1kpc) and medium-size symmetric objects (MSOs; $LS=1-10$ kpc) have been discovered 
by several authors (e.g., Wikinson et al. 1994; Fanti et al. 1995; 
Readhead et al. 1996). 
Among these compact radio sources, the bright objects show the FR I\hspace{-.1em}I-like morphology along with hot spots (reverse shocks), 
which are considered to be a signature of supersonic 
expansions (Owsianik, Conway \& Polatidis 1998; Taylor et al. 2000; Tschager et al. 2000; Giroletti et al. 2003; Polatidis \& Conway 2003; Gugliucci et al. 2005; Nagai et al. 2006; Gugliucci et al. 2007; Luo et al. 2007). 
In contrast, the faint compact radio sources do not show hot spots 
(e.g., Kunert-Bajraszewska et al. 2005; Giroletti 2007) 
and some of them show the FR I-like morphology (Giroletti 2007). 
Since the age of CSOs and MSOs is less than $\sim 10^{5}\,{\rm yr}$, 
they are recognized as newly born young AGN jets (e.g., Fanti et al. 1995; 
Readhead et al. 1996, O'Dea \& Baum 1997; Stanghellini et al. 1998; Snellen 
et al. 2000; Dallacasa et al. 2000; Orienti et al. 2007). 
Thus, CSOs and MSOs are adequate targets to examine the disk-jet connection in AGNs 
since the time lag between the optical and radio activity is relatively small 
($\Delta t({\rm AGNs}) \lesssim 10^{5-6}\, {\rm yr}$). 
In this paper, we will use bright CSOs and MSOs (hereafter 
powerful young radio-loud [YRL] AGNs) to investigate the disk-jet connection 
in AGNs.

In order to probe whether the formation of AGN jets requires any 
specific accretion conditions, it is essential to determine   
the physical states of the accretion disk of powerful YRL AGNs. 
To this end, it can be useful to examine a systematic difference in the 
SEDs produced by accretion disks of powerful YRL AGNs and radio-quiet (RQ) AGNs. 
However, it is challenging to derive the intrinsic SEDs of powerful YRL 
AGNs from the hard X-ray observations, because of the contamination 
of AGN jets and the obscuration of the accretion disk by the dusty torus. 
Alternatively, the optical emission-line diagnostics of 
narrow-line region (NLR) can be used to distinguish the physical 
states of the accretion disk, since the NLR is thought to be 
photoionized by the nuclear radiation for RQ Seyfert galaxies 
(e.g., Yee 1980; Shuder \& Osterbrock 1981; Evans et al. 1999) and 
even for extended radio galaxies (e.g., Villar-Martin et al. 1997; Nagao et al. 2006).
However, very little is currently known about the optical 
emission-line properties of powerful YRL AGNs. 
By examining the difference of optical line ratios between powerful YRL AGNs and 
the extended radio galaxies, Morganti et al. (1997) found that powerful YRL AGNs 
show the characteristics very similar to that of extended radio galaxies. 
However, it is not clear how the physical state of accretion disks 
(especially SEDs of accretion disks) is different between the powerful YRL AGNs 
and the RQ AGNs. Answering this question will provide crucial information on 
the disk-jet connection in AGNs. 

The main goal of this paper is to investigate whether there is the difference 
in SEDs between the powerful YRL AGNs and Seyfert 2 galaxies 
(i.e., RQ AGNs), by utilizing the optical emission-line diagnostics of NLRs. 
We collected samples of powerful YRL AGNs and RQ AGNs (\S 2), and generated
various photoionization models with different input SEDs (\S 3). In \S 4, we compared
the observational data with models. On the basis of our findings, 
Discussion on accretion conditions for the formation of powerful AGN jets is 
presented in \S5, and Conclusions follow in \S 6.

\section{Data compilation}
We focus on the properties of ionized gas in NLRs for powerful YRL 
AGNs to investigate the disk-jet connection.
To reduce possible inclination effects, it is important to compare AGNs with
the same optical type, since the optical narrow emission line ratios 
are often affected by the geometrical effects (e.g., Murayama \& Taniguchi 
1998; Schmitt 1998; Nagao, Murayama, \& Taniguchi 2000; Nagao, Murayama, 
\& Taniguchi 2001a). 
Out of 219 radio galaxies available in the literature, we compiled 32 YRL AGNs 
(20 narrow-line AGNs and 12 broad-line AGNs) with the optical spectra, from which 
the stellar continuum is subtracted, and high radio luminosity, $\log P_{\rm 5GHz} > 25.0\,{\rm W/Hz}$, using papers by Pearson \& Readhead (1988), 
Tadhunter et al. (1993) and O'Dea (1998). 
Since only few broad-line YRL AGNs have the detected emission lines 
(e.g., [O{\sc i}]$\lambda$6300 line, [O{\sc ii}]$\lambda$3727 and 
[O{\sc ii}]$\lambda$5007) which are essential for our study,
we finally selected 20 {\it narrow-line} YRL AGNs with these lines. 
In other words, the YRL AGN sample consists only of radio galaxies. 
The properties of powerful YRL AGNs are presented in Table 1. 
The mean projected linear size of selected radio sources is $3.2\,{\rm kpc}$, 
which is small enough to discuss the disk-jets connection in AGNs (see $\S 1$). The average radio power at 5 GHz, $\log P_{\rm 5GHz}$ is $\approx 27.0\,{\rm W/Hz}$, 
comparable to powerful extended radio galaxies (Laing, Riley \& Longair 1983). 
Note that the radio power of the faint YRL AGNs, excluded from our sample,
is $\log P_{\rm 5GHz}\leq 25.0\,{\rm W/Hz}$ (e.g., Giroletti 2007). 
We use only the forbidden lines to avoid the difficulty 
of separating the broad component and the narrow component of 
the recombination lines (e.g., ${\rm H}\alpha$ and ${\rm H}\beta$ lines). 
Then, we compiled the flux ratios of forbidden emission-lines from the literature (Table 2).
The typical errors of each line are $\approx 10-30\%$. 

To compare with narrow-line YRL AGNs, we selected a sample of RQ Seyfert 2 galaxies from
the Sloan Digital Sky Survey (SDSS; York et al. 2000) archive, 
using the emission-line flux catalog made by the 
MPA/JHU group (Kauffmann et al. 2003). 
We selected 624 RQ Seyferts, of which First Images of the Radio Sky 
at Twenty cm (FIRST) flux is lower than 1 mJy and 
optical emission-line fluxes are well-measured from high S/N ($> 10$) 
SDSS spectra.

In this paper, we do not evaluate the amount of extinction of 
host galaxies based on the Balmer decrement method (e.g., Osterbrock 1989), 
since for only few sources in the powerful YRL AGN sample detected were both 
the narrow H$\alpha$ and H$\beta$ lines. 
The data presented in Table 2 are not corrected for Galactic extinction. 
The effect of dust extinction on our results will be discussed in \S4.1. 

\section{Photoionization model}
To characterize SEDs produced by the accretion disk in powerful YRL AGNs,
we will compare emission line properties with photoionization models.
We will examine whether there is a systematic difference 
in SEDs from accretion disks between the powerful YRL AGNs 
and the RQ AGNs. The continuum radiation from the accretion disk photoionizes the NLR, hence, the difference in the SED can be related to the difference 
in the physical states of the NLR. Then, we compare the observed flux ratios 
with photoionization model calculations, using the publicly available code CLOUDY, version 06.02c (Ferland 1998).

The method of our calculations is based on
solving the equations of statistical equilibrium, the ionization and thermal equilibrium through gas clouds as a function of depth. 
Then, we obtain the ionization structure of gas in NLRs and 
the volume emissivity of the forbidden emission lines 
(i.e., [O{\sc i}]$\lambda$6300, [O{\sc ii}]$\lambda$3727, [S{\sc ii}]$\lambda$6717+6731, [O{\sc iii}]$\lambda$ 5007 and [Ne{\sc iii}]$\lambda$3869). 
Here we briefly describe the adopted parameterizations. 
We assume uniform density gas clouds with a plane-parallel geometry. 
The parameters for the calculations are (i) the hydrogen density 
($n_{\rm H}$); (ii) the ionization parameter ($U=Q({\rm H})
/4\pi c r^{2}n_{\rm H}$, 
where $Q({\rm H})$ is the number of the ionizing photons emitted by AGNs 
per second), which is defined as the ratio of the ionizing density 
to the hydrogen density; 
(iii) the chemical composition of the gas; (iv) the shape of SEDs of 
the input continuum radiation. While the gas metallicity is assumed to be solar (Grevesse \& Sauval 1998; Holweger 2001), we carry out a series of model runs covering various density and ionization parameters:
$10^{2}\,{\rm cm}^{-3}\leq n_{\rm H} \leq 10^{5}\,{\rm cm}^{-3}$ and $10^{-4.0} \leq U \leq 10^{-1.5}$. 

For the input continuum radiation, we consider two types of SEDs.
One is a typical SED produced by a standard accretion disk (Shakura \& Sunyaev 1973), which is characterized by a strong big blue bump (BBB) 
in the wavelength range of UV to soft X-ray. 
As a template of this type of SED (hereafter, SED with BBB), 
we adopt the empirical expression using the following function (see Ferland 1997; Nagao et al. 2001b): 
\begin{equation}
f_{\rm \nu}=\nu^{\alpha_{\rm UV}}{\rm exp}\left(-\frac{h\nu}
{kT_{\rm BB}}\right){\rm exp}\left(-\frac{kT_{\rm IR}}{h\nu}\right)+
a\nu^{\alpha_{\rm x}}, 
\end{equation}
where the infrared cutoff of the BBB component, $kT_{\rm IR}=0.01
\,{\rm ryd}$ ($1\,{\rm ryd}=13.6\,{\rm eV}$), the characteristic temperature 
of the BBB, $T_{\rm BB}=4.9\times 10^{5}\, {\rm K}$, 
the slope of the low-energy side of the BBB, $\alpha_{\rm UV}=-0.5$, 
the UV to X-ray spectral index, $\alpha_{\rm ox}=-1.35$ and the slope of 
the X-ray power-law continuum, $\alpha_{\rm x}=-0.85$ which are typical 
values of Seyfert galaxies (see Nagao et al. 2001b for more details). 
Since the characteristic temperature of the BBB in some extreme AGNs, e.g.,
narrow-line Seyfert 1 galaxies, can reach at $T_{\rm BB}\approx 10^{6}\, {\rm K}$, 
we will discuss the effects of changing $T_{\rm BB}$ on the line flux ratios. Note that the parameter $a$ in eq. (1) is determined from the adopted value of $\alpha_{\rm ox}$. 
The last term in equation (1) is set to zero below 1.36 eV,
and the continuum is assumed to fall off as $\nu^{-3}$ above 100 keV.

The other kind of SED is a typical SED produced by a RIAF,
which is expected to show a power-law with no strong BBB as shown by 
various works (e.g., Narayan \& Yi 1995; Chen, Abramowicz, \& Lasota 1997; 
Narayan, Kato, \& Honma 1997; Manmoto, Mineshige \& Kusunose 1997; 
Kurpiewski \& Jaroszynski 1999; Quataert \& Narayan 1999; Manmoto 2000; 
Kino, Kaburaki, \& Yamazaki 2000; Ohsuga, Kato \& Mineshige 2005; 
Yuan, Ma \& Narayan 2008). 
We adopt here the SED model presented by Kurpiewski \& Jaroszynski (1999) 
as the template for SEDs of RIAF (hereafter SED without BBB). 
This SED can be expressed by a single power-law continuum in the range 
of $10^{12}$ Hz to $10^{20}$ Hz,
with the exponential cutoffs at $10^{-4}$ ryd and $10^{4}$ ryd. 
We choose the photon index of $\alpha_{\rm PL} =0.89$ ($f_{\rm \nu}=\nu^{-\alpha_{\rm PL}}$), 
as predicted for the case of a non-rotating black hole.
Since there are various shapes of the calculated SED for a RIAF 
(i.e., various $\alpha_{\rm PL}$), we will check the effect of altering 
$\alpha_{\rm PL}$ in the wide range of $0.78 < \alpha_{\rm PL} < 1.0 $.

Using these two SED templates for a standard disk and  RIAF, respectively 
(see Figure 3 in Nagao et al. 2002), 
we calculate emission line flux ratios using CLOUDY. 
These simple SED templates can be used to elucidate the effect of 
the presence or absence of the BBB component, which is the main 
difference in the SED between a standard disk and RIAF. 
Note that these template might be too simple to generate accurate 
narrow-emission line fluxes in AGNs. 
The calculations are stopped when the temperature falls below 3000 K, 
where the gas does not contribute significantly to the 
observed optical emission line spectra.

\section{Results}
In this section, we compare the photoionization model calculations with
observed emission line properties of powerful YRL AGNs and Seyfert 2 galaxies.

\subsection{Comparison with models with/without BBB}
The emission line ratio, [O{\sc i}]$\lambda$6300/[O{\sc iii}]$\lambda$5007,
of the NLR can be used to characterize the SED of accretion disk since 
the line ratio depends on the shape of SED of the photoionizing source.
The harder spectra (without BBB) of optically thin disks create larger 
ratios [O{\sc i}]$\lambda$6300/[O{\sc iii}]$\lambda$5007 due to the following 
reasons.
The mean free path of ionizing photons at higher energies becomes longer, 
since the cross section for ionization of hydrogen ($\sigma$(H)) is given by 
$\sigma\,({\rm H})\propto (\nu/\nu_{\rm L})^{-3}$ for $\nu > \nu_{\rm L}$, where $\nu_{\rm L}$ is the frequency of the Lyman limit. 
With a flat power-law spectrum ($\alpha_{\rm PL} \lesssim 1$), 
a relatively large number of photons are available at higher energies 
even after the photons near the Lyman limit are all absorbed 
around the ionization front. 
Then, the higher energy photons can reach at larger radii from the 
ionization source, hence, the extended partially ionized region 
can be formed. Furthermore, the low-ionization emission 
arises in the extended partially ionized region rather than the 
fully ionized region where the abundance of neutral oxygen is negligible. 
Therefore, the volume emissivity of low-ionization emission lines 
(i.e., [O{\sc i}]$\lambda 6300$) can be higher at larger radii from 
the ionization sources, compared to that of high-ionization emission lines 
(i.e., [O{\sc iii}]$\lambda 5007$) (see Figure 5 of Nagao et al. 2002). 
Note that the ionization potential of neutral oxygen is nearly identical to 
that of hydrogen.

Figure 1 represents emission line ratios ([O{\sc i}]$\lambda$6300/[O{\sc iii}]$
\\ \lambda$5007 versus 
[O{\sc ii}]$\lambda$3727/[O{\sc iii}]$\lambda$5007) from our model calculations and observations. 
The horizontal axis ([O{\sc ii}]$\lambda$3727/[O{\sc iii}]$\lambda$5007) indicates the difference of $n_{\rm H}$ because the critical density of [O{\sc ii}]$\lambda$3727 ($n_{\rm H}=4.5\times 10^{3}\,{\rm cm}^{-3}$) is far smaller than that of [O{\sc iii}]$\lambda$5007 ($n_{\rm H}=7.0\times 10^{5}\,{\rm cm}^{-3}$). 
This set is free from the chemical abundance effect. 
As shown in Figure 1, powerful YRL AGNs have relatively higher [O{\sc i}]$\lambda$6300/[O{\sc iii}]$\lambda$5007
than Seyfert 2 galaxies, implying that there is a difference in the shape of ionizing continuum.
The observed line ratios were not corrected for both the Galactic extinction 
or internal extinction of host galaxies.
Instead, we indicate by the arrow (upper left in Figure 1) the effect of 
dust extinction 
when the dust-extinction correction of $A_{\rm v}=1.0$ mag is applied, 
assuming a standard Galactic reddening law (Cardelli et al. 1989). 
For type 1 AGNs, the dust reddening of NLR emission is small ($A_{\rm v}< 1.0$ mag;
e.g., Rodriguez-Ardila et al. 2000). 
The dust-extinction of NLRs for type 2 AGNs is 
approximately 1.0 mag larger than that for type 1 AGNs
(e.g., De Zotti \& Gaskell 1985; Dahari \& De Robertis 1988).

The models using SED without BBB predict higher [O{\sc i}]$\lambda$6300/[O{\sc iii}]$\lambda$5007 ratio than the models using SED with BBB. 
The trend of the model prediction coincides with the observations. 
The line ratios of Seyfert 2 galaxies are consistent with models adopting 
the SED with BBB in the range of $10^{-3.5} < U < 10^{-3.0}$ 
and $10^{4.0}\,{\rm cm}^{-3} < n_{\rm H} < 10^{5.0}\,{\rm cm}^{-3}$, while 
the line ratios of powerful YRL AGNs can be explained 
by the model adopting the SED without BBB. 
Note that the derived $n_{\rm H}$ may not reflect the actual number density 
of narrow-line clouds because of the dust extinction effects. 
However, the direction of extinction correction is 
perpendicular to the direction of the difference between 
powerful YRL AGNs and Seyfert 2 galaxies, indicating that  
the observed difference is not due to the difference in the 
degree of the dust extinction between powerful YRL AGNs and 
Seyfert 2 galaxies. 

In Figure 2, we present the frequency distributions of $\log($[O{\sc i}]$\lambda$6300/[O{\sc iii}]$\lambda$5007) 
for the powerful YRL AGNs (blue) and Seyfert 2 galaxies (red).
The mean value of $\log($[O{\sc i}]$\lambda$6300/[O{\sc iii}]$\lambda$5007) 
for powerful YRL AGNs is $-0.36\pm0.26$, while that of $\log($[O{\sc i}]$
\lambda$6300/[O{\sc iii}]$\lambda$5007) for Seyfert 2 galaxies is $-1.17\pm0.05$.
The Kolmogorov-Smirnov (K-S) test indicates $10^{-4}$ probability that the observed 
frequency distributions of [O{\sc i}]$\lambda$6300/[O{\sc iii}]$\lambda$5007 
of powerful YRL AGNs and Seyfert 2 galaxies originate from the same underlying 
population, suggesting that the emission line ratios of 
[O{\sc i}]$\lambda$6300/[O{\sc iii}]$\lambda$5007 for powerful YRL AGNs 
are statistically larger than those of Seyfert 2 galaxies.  

In figure 3, we compare the observations with the model predictions in other 
diagnostic diagrams, i.e., [O{\sc i}]$\lambda$6300/[O{\sc iii}]$\lambda$5007 
versus [S{\sc ii}]$\lambda$6717+6731/[O{\sc iii}]$\lambda$5007 and 
[O{\sc i}]$\lambda$6300/[O{\sc iii}]$\lambda$5007 versus 
[Ne {\sc iii}]$\lambda$3869/[O{\sc ii}]$\lambda$3727.  
The physical meaning of these diagrams is same as Figure 1. 
Similar to the trend in Figure 1, powerful YRL AGNs and Seyfert 2 galaxies 
are well separated in these diagnostic diagrams, suggesting that 
the different line ratios of powerful YRL AGNs can be explained 
by models with the photoionizing SED without BBB.
However, one may doubt that the difference of SED is not a unique 
solution to explain the observed differences. 
Thus, we discuss the other possibilities for high [O{\sc i}]
$\lambda$6300/[O{\sc iii}]$\lambda$5007 ratios in $\S 5.1$.

\subsection{Dependence on the model parameters}
Here, we examine the dependence of emission line ratios on the model input parameters, $T_{\rm BB}$ and $\alpha_{\rm PL}$. 
Figure 4 shows the optical emission line ratios ([O{\sc i}]$\lambda 6300$/
[O{\sc iii}]$\lambda 5007$ and [O{\sc ii}]$\lambda3727$/[O{\sc iii}]
$\lambda 5007$) as a function of $\alpha_{\rm PL}$ (left panel) and $T_{\rm BB}$ (right panel), assuming $n_{\rm H}=10^{4}\,{\rm cm}^{-3}$. 
We find that at given $U$ the variations of $T_{\rm BB}$ and 
$\alpha_{\rm PL}$ does not change significantly the line ratios of 
[O{\sc i}]$\lambda 6300$/[O{\sc iii}]$\lambda 5007$, compared to the effect
of the difference in SEDs of ionizing radiation.
Since the change of emission line ratios due to $\alpha_{\rm PL}$ or $T_{\rm BB}$ is relatively small, 
the effect of changing $T_{\rm BB}$ and $\alpha_{\rm PL}$ 
cannot account for the observed difference between powerful YRL AGNs and 
Seyfert 2 galaxies. 
The effect of both $T_{\rm BB}$ and $\alpha$ does not also change 
the results shown in Figure 3. 
Therefore, the difference in the observed emission line ratios 
in Figure 1 and Figure 3 is consistent with the idea that the 
NLR of powerful YRL AGNs is photoionized by SEDs without BBB, while 
that of the RQ AGNs is photoionized by SEDs with BBB.

\subsection{[O{\sc iii}]$\lambda$5007/[O{\sc iii}]$\lambda$4363}
In Figure 5, we investigate [O{\sc iii}]$\lambda$5007/[O{\sc iii}]$\lambda$4363 line ratios, which is one of the gas temperature indicators (Osterbrock 1989).
The line ratio ([O{\sc iii}]$\lambda$5007/[O{\sc iii}]$\lambda$4363) of 
powerful YRL AGNs is smaller than that of Seyfert 2 galaxies.
The mean value of $\log($[O{\sc iii}]$\lambda$5007/[O{\sc iii}]$\lambda$4363) 
for powerful YRL AGNs is $1.50\pm0.29$, while that for Seyfert 2 galaxies 
is $2.03\pm0.16$. 
The frequency distributions of these line ratios for the powerful 
YRL AGNs (blue) and Seyfert 2 galaxies (red) are presented in Figure 6. 
The K-S test indicates $10^{-4}$ probability that the observed 
frequency distributions of [O{\sc iii}]$\lambda$5007/[O{\sc iii}]$\lambda$4363 
of YRL AGNs and Seyfert 2 galaxies originate in the same underlying population. We will discuss potential physical reasons for smaller [O{\sc iii}]$\lambda$5007/[O{\sc iii}]$\lambda$4363 of powerful YRL AGNs in $\S 5.3$.  

\section{Discussion} 
\subsection{Other possibilities for high [O{\sc i}]$\lambda$6300/[O{\sc iii}]$\lambda$5007 ratios}
We here consider three other plausible possibilities which may explain the 
observed difference in line ratios.

i) Shock ionization. We consider the shock ionization due to the interaction between AGN jets and the clouds in NLRs. The strong low-ionization emission lines 
such [O{\sc i}]$\lambda 6300$ causes by the shock-heated gas 
(e.g., Mouri, Kawara, \& Taniguchi 2000 and references therein) because shocks generate a large partially ionized region in the gas. 
To investigate whether or not the shock heating by AGN jets is 
responsible for the difference in the narrow emission-line ratio 
between powerful YRL AGNs and RQ AGNs, 
we compare the observed line ratios with shock models in Figure 7. 
The adopted shock models with/without considering the effect of 
precursor \footnote{This effect is that the unshocked gas is ionized 
by the radiation emitted by the hot shocked gas.} 
are presented by Dopita \& Sutherland (1995). 
The shock models predict too small flux ratio of 
[O{\sc i}]$\lambda$6300/[O{\sc iii}]$\lambda$5007, too small flux ratio of 
[Ne{\sc iii}]$\lambda$3869/[O{\sc ii}]$\lambda$3727 and 
too large [O{\sc ii}]$\lambda$3727/[O{\sc iii}]$\lambda$5007, 
compared with the observed line ratios of powerful YRL AGNs. 
To explain all observations (Figure 1 and Figure 3) by shock models, 
a fine-tunning of model parameters should be required. 
Thus, these results suggest that the difference in the narrow-line flux ratios 
of powerful YRL AGNs are  not mainly caused 
by the shock-heated gas in NLRs. 
However, we should mention that for one of YRL AGNs in our sample 
(3C 303.1) the shock ionization cannot be negligible (e.g., 
Labiano 2005). 
Note that the NLR properties of 
powerful YRL AGNs can be explained by the photoionization model 
(on average), in spite of the signature of the strong kinematic interaction 
with surrounding gas (e.g., Axon et al. 2000; O'Dea et al. 2002; Holt et al. 2008; Labiano et al. 2008). 

ii) Gas density. The systematic difference in the gas density 
in NLR may cause the different line ratios, because the critical density of 
[O{\sc i}]$\lambda$6300 transition ($n_{\rm H}=1.8\times 10^{6}\,{\rm cm}^{-3}$) is larger than that of [O{\sc iii}]$\lambda$5007 transition 
($n_{\rm H}=7.0\times 10^{5}\,{\rm cm}^{-3}$). 
This possibility can be easily rejected as shown in Figure 1 and Figure 3. 
At given ionization parameter $U$, 
the ratio of [O{\sc i}]$\lambda$6300/[O{\sc iii}]$\lambda$5007 does not depend 
on the gas density, if we employ only the models with SEDs with BBB. 
This is because both critical densities of [O{\sc i}]$\lambda$6300 transition 
and [O{\sc iii}]$\lambda$5007 transition are higher than the 
typical gas density of NLRs in AGNs ($10^{2.0}\,{\rm cm}^{-3} \leq  n_{\rm H} 
\leq 10^{5.0}{\rm cm}^{-3}$). 
This indicates that the observed difference in the flux ratio of 
[O{\sc i}]$\lambda$6300/[O{\sc iii}]$\lambda$5007 is not caused only 
by a systematic difference in the gas density of NLRs.

iii) Ionization parameter. We consider the difference in the ionization parameter 
$U$ between powerful YRL AGNs and RQ AGNs. 
Since the effect of the SED difference and changing the ionization 
parameter (i.e., the ratio of the ionizing photon density to 
the hydrogen density) are often degenerated (see Nagao et al. 2002), 
a systematic difference in the ionization parameter of 
gas in NLRs between powerful YRL AGNs and RQ AGNs can  
explain the difference in the emission line ratios [O{\sc i}]$\lambda 6300$/[O{\sc iii}]$\lambda$5007 when we fix $n_{\rm H}$. 
As shown in Figure 1 and Figure 3 (bottom), we find that only different $U$ 
cannot explain the difference of line ratios between powerful YRL AGNs and RQ AGNs. 
However, it is possible that the combination of $n_{\rm H}$ and $U$ 
can reproduce the difference between them even if we assume the SEDs with BBB. 
If the NLR of powerful YRL AGNs has generally higher $n_{\rm H}$ and lower $U$ 
than that of Seyfert 2 galaxies, the observed difference can be explained 
without invoking different shape of photoionizing SEDs.
This possibility cannot be completely ruled out from Figure 1 and Figure 3, 
however, a special fine-tunning of model parameters is required to generate
the observed line ratios of powerful YRL AGNs.

The degeneracy of the effect of SED shape and of the ionization parameter is a limitation
in interpreting these results. For the future studies,
the diagnostic diagram using the flux ratio of [Ar{\sc iii}]$\lambda$7136/[O{\sc iii}]$\lambda$5007 (e.g., [Ar{\sc iii}]$\lambda$7136/[O{\sc iii}]$\lambda$5007 versus [O{\sc ii}]$\lambda$3727/[O{\sc iii}]$\lambda$5007 and [Ar{\sc iii}]$\lambda$7136/[O{\sc iii}]$\lambda$5007 versus [Ne{\sc iii}]$\lambda$3727/[O{\sc iii}]$\lambda$5007) can be used to determine the ionization parameter, 
since [Ar{\sc iii}]$\lambda$7136/[O{\sc iii}]$\lambda$5007 
is almost independent of the shape of SED in the range of $U\lesssim
10^{-2.5}$ (Nagao et al. 2002). Thus, it can be better constrained 
whether the SEDs of YRL AGNs are different from those of Seyfert 2 galaxies.
Unfortunately, these lines are not currently available for our sample of YRL AGNs.

\subsection{Disk-jet connection} 
In $\S 4$, we have shown that the powerful YRL AGNs 
favor SEDs without a strong BBB such as the one expected for RIAFs. 
On the other hand, it is well-known that RIAFs are radiatively 
inefficient and generate relatively  lower Eddington ratios
(e.g., Narayna \& Yi 1995; Kato, Fukue \& Mineshige 1998). 

The adovection-dominated solution exists when the mass accretion rate is 
relatively low as $\dot{M}\leq \dot{M}_{\rm max}$. 
If we assume a fiducial efficiency is $\approx 10\%$ 
(Frank, King \& Raine 1992), the maximum accretion rate is given 
by $\dot{M}_{\rm max}\approx 3\alpha^{2}L_{\rm Edd}/c^{2}$ 
at $r < 100r_{\rm g}$, where $r_{\rm g}$ is the Schwarzshild radius of 
the central black hole, $\alpha$ is the 
$\alpha$-viscosity (Shakura \& Sunyaev 1973) and $L_{\rm Edd}$ is the 
Eddington luminosity corresponding to the Thomson scattering. 
The energy release from a RIAF ($L_{\rm RIAF}$) can be obtained as 
$L_{\rm RIAF}\approx 0.1 (\dot{M}/\dot{M}_{\rm max})\dot{M}c^{2}$ 
(e.g., Kato, Fukue \& Mineshige 1998). According to three-dimensional 
magneto-hydrodynamic simulations, $\alpha$ is $\alpha \lesssim 0.1$ 
(e.g., Balbus \& Hawley 1991; Machida et al. 2000; Machida \& Matsumoto 2003). 
We here adopt the conservative value as $\alpha=0.1$ 
and then the maximum luminosity of a RIAF 
at $\dot{M}=\dot{M}_{\rm max}$ is given as $L_{\rm RIAF, max}
\approx 10^{-3}L_{\rm Edd}$.

Thus, it is necessary to consider whether RIAFs are able to explain the 
relatively high [O{\sc iii}]$\lambda$5007 luminosity (hence, high bolometric
luminosity) of powerful YRL AGNs.
First, we estimated the bolometric luminosity of powerful YRL AGNs 
based on the [O{\sc iii}] luminosity. 
The [O{\sc iii}] luminosity of 16 powerful YRL AGNs were collected from
Gelderman \& Whittle 1994 and Lawrence et al. 1996 (see Table 1). 
The average [O{\sc iii}] luminosity is $\bar{L}_{\mbox{\tiny[O{\sc iii}]}}=10^{42.24\pm 1.25}\,{\rm erg/s}$. 
To evaluate the bolometric luminosity, we adopt the bolometric 
correction of low luminosity AGNs (LLAGNs) whose SEDs can be well 
expressed by a RIAF model because of the lack of BBB (e.g., Ho 1999), 
i.e.,  $L_{\rm bol}\approx 250\, L_{\mbox{\tiny[O{\sc iii}]}}$ 
with variance of 0.9 dex (Panesa et al. 2006)
\footnote{$\log(L_{2-10\,{\rm keV}}/L_{\mbox{\tiny[O{\sc iii}]}})=
1.5\pm 0.5$ and $\log(L_{\rm bol}/L_{2-10\,{\rm keV}})=0.9\pm 0.4$}. 
Then, the average bolometric luminosity of YRL AGNs, 
$\bar{L}_{\rm bol, obs} \sim 10^{45}\, {\rm erg/s}$, is required to produce 
the observed [O{\sc iii}] line luminosity. This suggests that 
$\bar{L}_{\rm bol, obs}$ is $\sim 10^{-2}\, L_{\rm Edd}$ 
for the massive BHs ($M_{\rm BH}=10^{9}M_{\odot}$). 
Hence, it seems difficult to explain the bolometric luminosity of 
powerful YRL AGNs with the RIAF model, because of $L_{\rm RIAF, max}
\lesssim 0.1\bar{L}_{\rm bol, obs}$. 
The combination of emission line ratios and the bolometric luminosity 
estimates implies that the physical nature of the 
accretion disk of YRL AGNs cannot be explained by either the classical steady 
disk, i.e., a standard disk or a RIAF. 
It might imply that the relativistic jet occur in a non steady accretion disk.
 
Interestingly, the physical states of accretion disks 
in some XRBs (e.g., GX 339$-$4) associated with relativistic jets 
are similar to those of YRL AGNs (Homan et al. 2005; Remillard 2005),
suggesting  the similarity of disk-jet connection in XRBs and AGNs.
This similarity may indicate that the formation of relativistic jets 
requires luminous accretion disk with harder ionizing SED. 
However, we note that the bolometric luminosity estimated from [OIII] 
luminosity is highly uncertain for both luminous AGNs and LLAGNs without 
BBB.
Thus, it is crucial to reduce the uncertainty of bolometric luminosity 
by evaluating the bolometric correction using a large sample of broad-line
YRL AGNs, which are left in our future works.

\subsection{Obscuring structure in powerful YRL AGNs}
We discuss here why the line ratios, $\log($[O{\sc iii}]$\lambda$5007/[O{\sc iii}]$\\ \lambda$4363), of powerful YRL AGNs is smaller than that of Seyfert 2 
galaxies. As we investigated in \S 4, several optical narrow line emission 
ratios ([O{\sc i}]$\lambda$6300/[O{\sc iii}]$\lambda$5007, 
[O{\sc ii}]$\lambda$3727/[O{\sc iii}]$\lambda$5007, [S{\sc ii}]$\lambda$6717+6731/[O{\sc iii}]$\lambda$5007 and [Ne{\sc iii}]$\lambda$3869/[O{\sc ii}]$\lambda$3727) can be well explained by single-zone photoionization models. 
However, it has been well-known that any one-zone models overpredicts the flux ratio of [O{\sc iii}]
$\lambda$5007/[O{\sc iii}]$\lambda$4363 (e.g., Koski \& Osterbrock 1976; Heckman 1980; Ferland \& Netzer 1983; Rose \& Cecil 1983; Keel \& Miller 1983; 
Rose \& Tripicco 1984).

By taking high-density gas clouds into account in NLRs, 
Nagao et al. (2001c) reported that photoionization models can explain low 
emission-line flux ratios of [O{\sc iii}]$\lambda$5007/[O{\sc iii}]$
\lambda$4363 $\sim 10$, without including shock effects. 
This idea is attributed to the fact that the critical density of 
[O{\sc iii}]$\lambda$4363 ($n_{\rm H}=3.3\times 10^{7}\,{\rm cm}^{-3}$) 
is higher than that of [O{\sc iii}]$\lambda$5007 ($n_{\rm H}=
7.0\times 10^{5}\,{\rm cm}^{-3}$). 
In addition, they suggested that the [O{\sc iii}]$\lambda$4363 originates 
in the dense gas clouds which are located at the inner regions 
compared with [O{\sc iii}]$\lambda$5007, hence, a large fraction of gas clouds 
emitting [O{\sc iii}]$\lambda$4363 line are obscured in type 2 AGNs.
This idea is supported by the observations that the line ratios of 
[O{\sc iii}]$\lambda$5007/[O{\sc iii}]$\lambda$4363 
of Seyfert 2 galaxies are larger than those of Seyfert 1 galaxies 
(Nagao et al. 2001c). 
If this is the case, the difference in [O{\sc iii}]$\lambda 5007$/[O{\sc iii}]$\lambda$4363 may reflect the difference in the structure of duty torus, i.e., 
the difference of the geometrical thickness. 
We should emphasize that the difference in Figure 5 (also Figure 6) is not related to the inclination effect, because we compare the same optical types of AGNs, i.e., powerful narrow-line YRL AGNs and Seyfert 2 galaxies. 
Since the ratio [O{\sc iii}]$\lambda 5007$/[O{\sc iii}]$\lambda$4363 of 
powerful YRL AGNs is lower than that of Seyfert 2 galaxies, 
the obscuring torus of powerful YRL AGNs may be 
thinner than that of Seyfert 2 galaxies (see Figure 8). 
According to a recent coevolution model of the BH growth and the circumnuclear disk (Kawakatu \& Wada 2008; see also Wada \& Norman 2002), the mass accretion 
rate onto a central BH driven by the turbulent viscosity is higher 
as the scale height of circumnuclear disk increases. 
This is because the angular momentum transfer due to the turbulence is proportional to the the scale height.  
To test our scenario, it is crucial to examine the difference 
in the structure of obscuring torus, 
using Atacama Large Millimeter/submillimeter Array (ALMA).

In summary, our findings indicate that the powerful 
YRL AGNs favor the accretion disk without BBB, surrounded by 
the geometrically thin obscuring torus,
while Seyfert 2 galaxies prefer the accretion disk with BBB, surrounded 
by the geometrically thick torus (see Figure 8). 
These results suggest that the formation of the powerful AGN jet is
related to the physical states of accretion disks and structure of obscuring 
materials around the accretion disk. 

\section{Conclusions}
We examine the optical narrow emission-line flux ratios of NLRs, 
in order to determine whether the formation of powerful 
AGN jets requires specific accretion conditions. 
A sample of powerful YRL AGNs, which are at the early stage of 
the jet activity, and radio-quiet Seyfert 2 galaxies are selected. 
We summarize our main results.

\begin{enumerate}
\item
Powerful YRL AGNs exhibit larger flux ratios of 
[O{\sc i}]$\lambda$6300/[O{\sc iii}]$\lambda$5007 than RQ Seyfert 2 galaxies. 
By comparing the observed line ratios with the photoionization model calculations,
we find that the difference between powerful YRL AGNs and Seyfert 2 galaxies 
can be well understood by the difference in SEDs of ionizing radiation 
from the accretion disk. 
The powerful YRL AGNs favor SED without a strong BBB such as the ADAF, 
while the line ratios of Seyfert 2 galaxies are consistent with the models 
adopting SED with a strong BBB (a geometrically thin, optically thick disk). 
In contrast, the observed difference in the line ratios is difficult to explain 
by the difference in the contribution of shocks caused by the AGN jets. 

\item
The line ratio [O{\sc iii}]$\lambda 5007$/[O{\sc iii}]$
\lambda$4363 of powerful YRL AGNs is systematically smaller than that of RQ 
AGNs. If we adopt a scenario that [O{\sc iii}]$\lambda$4363 originates 
in the dense gas clouds, which are located at the inner regions compared with 
[O{\sc iii}]$\lambda$5007, it is inferred that the obscuring torus around powerful 
YRL AGNs would be thinner than that around Seyfert galaxies. 

\end{enumerate}

\acknowledgments
We thank an anonymous referee for useful and helpful comments.
NK is financially supported 
by the Japan Society for the Promotion of Science (JSPS) through 
the JSPS Research Fellowship for Young Scientists. 
TN is financially supported through the Research Promotion Award 
of Ehime University. JW acknowledges the support provided by NASA through Hubble Fellowship grant HF-0642621
awarded by the Space Telescope Science Institute, which is operated by the Association of Universities for Research in Astronomy, Inc., for NASA, under contract NAS 5-26555.

\clearpage
\begin{table}[t]
\begin{center}
Table 1. Properties of powerful young radio-loud AGNs  \\[3mm]
{\scriptsize
\begin{tabular}{lccccc}
\hline \hline
Name & z & $\log{P_{\rm 5\,{\rm GHz}}}$ (W/Hz) & Size (kpc) & $\log{L_{\rm [OIII]}}$ (erg/s) & References \\  
(1) & (2) & (3) & (4) & (5) & (6) \\
\hline 
0023$-$26  &0.322 & 27.0 & 4.06 & --- &  1 \\
0316+413 (3C 84) &0.017 & 25.1 & $4.6\times 10^{-3}$ & 39.9 & 3,5 \\
0345+337 (3C 93.1) &0.244 & 26.0 & 4.06 &40.6 &  2,3 \\
0404+769 (4C 76.03) &0.598 & 27.4& 0.86 &41.9&  2,4 \\
0518+165 (3C 138) &0.760 & 27.8 & 3.94 &44.1 & 2 \\
0605+480 (3C 153) &0.277 & 25.7 & 13.1& 41.7 & 3,5   \\
0710+439 &0.518 & 27.1  & 0.14 & 42.4 & 2,4 \\
0954+658 &0.899 & 26.4 & $\sim 1$ & 43.1 & 4,6 \\
1031+567 &0.459 & 26.9  & 0.21 & 41.6 & 2,4 \\
1306$-$09  &0.464 & 27.1 & 2.72 & --- &  1 \\
1328+307 (3C 286) &0.849 & 28.2 & 22.0 &44.3 & 2,3 \\
1345+125 (4C 12.50) &0.122 & 26.0 & 0.16 & 43.0 & 2,3 \\
1358+624 (4C 62.22)&0.429 & 26.9 & 0.35 & 41.8 & 2 \\
1443+77  (3C 303.1)&0.267 & 26.0 & 6.22 &42.9 & 2,3 \\
1634+628 (3C 343) &0.988 & 27.7 & 1.45&--- & 2 \\
1637+626 (3C 343.1) &0.750 & 27.4 & 1.57 &43.3 &2,3 \\
1807+698 (3C 371)&0.050 & 25.0 & $\sim 1$ & 40.7 & 3,6 \\
1934$-$63  &0.183 & 26.7 & 0.17 & --- &  1 \\
2342+821 &0.735 & 27.4 & 1.17 & 43.1 & 2,4 \\
2352+495 &0.237 & 26.3 & 0.17 & 41.3 & 2,4 \\
\hline
\end{tabular}
}
\noindent
\end{center}
{\scriptsize Note.-All quantities are caluculated assuming 
$H_{\rm 0}=75\,{\rm km}\,{\rm s}^{-1}\,{\rm Mpc}^{-1}$ and 
$q_{0}=0$. Col. (1): Object name. Col. (2): redshift. Col. (3): 
log of radio power at 5 GHz. Col. (4) linear size of the radio jets. 
Col. (5): log of [OIII]$\lambda$5007 luminosity. \\ 
References.-(1) Tadhunter et al. 1993, (2) O'Dea 1998, 
(3) Gelderman \& Whittle 1994, (4) Lawrence et al. 1996, 
(5) Kawakatu, Nagai \& Kino 2008, (6) Lister et al. 2001,  
}
\end{table}

\clearpage
\begin{table}[t]
\begin{center}
Table 2. Emission-line ratios relative to H$\beta$ flux  \\[3mm]
{\scriptsize
\begin{tabular}{lcccccccc}
\hline \hline
Name & [OII] & [NeIII] & [OIII] & [OIII] & [OI] & [SII] & [SII] & References \\
--- & $\lambda 3727$ & $\lambda 3869$ & $\lambda 4363$ & $\lambda 5007$ 
& $\lambda 6300$ &  $\lambda 6716$ & $\lambda 6731$ &--- \\ 
(1) & (2) & (3) & (4) & (5) & (6) & (7) & (8) & (9) \\
\hline 
0023$-$26 &4.90  & 0.51  & 0.10 & 3.40  & 2.70 & 2.48 & 3.10 & 1  \\
0316+413 & 2.95 & 0.92 & 0.38 & 10.9 & 6.61 & 3.37 & 2.67 & 3 \\
0345+337 & ---  & ---  & ---  & 8.33 & 2.00 & --- & ---  &  2 \\
0404+769 & ---  & 0.33 & ---  & 3.66 & ---  & --- &  --- &  3 \\
0518+165 & ---  & ---  & ---  & 0.93 & ---  & --- &  --- &  2 \\
0605+480 & 5.25 & ---  & ---  & 4.74 & 2.26 & 3.17 & 2.71 & 3 \\
0710+439 & 2.54 & 0.44 & 0.28 & 6.10 & 1.33 &---  &  --- &  3 \\
0954+658 & 0.45 & 0.47 & 0.33 & 7.48 & ---  & ---  & ---  & 3 \\
1031+567 & 3.16 & 0.98 & ---  & 7.20 & 5.80 & --- &  --- &  3 \\
1306$-$09 & 4.64 & 0.60  & 0.06 &  3.57 & ---  & ---  & ---  &  1 \\
1328+307 & 0.48 & 0.57 & 0.39 & 4.34 & ---  & --- & ---  &  2 \\
1345+125 & 2.19 & 1.06 & 0.38 & 6.25 & 0.94& --- &  --- &  2 \\
1358+624 & 2.49 & 0.78 & 0.10& 8.69 & 5.61 &3.49 & 5.04 &  3 \\
1443+77  & 4.58  & 0.50 & ---  & 9.00  & ---  & --- & ---  &  4 \\
1634+628 & 2.06 & 1.08 & 0.18 &---   & ---  & --- & ---  &  3 \\
1637+626 & ---  &---   & ---  & 6.33 & ---  & --- & ---  &  2 \\
1807+698 & 1.10 & ---  &  --- & 3.66 & 2.34 & 1.06 & 2.01 & 3 \\
1934$-$63 & 3.12 &  1.44 & 0.26 & 7.21  & ---  & ---  & ---  &  1 \\
2342+821 & 0.94 & 0.59 & ---  & 9.98 & ---  & --- &  --- &  3 \\
2352+495 & 3.21 & 0.80 & ---  & 8.56 & 3.88 & 3.25&  3.01&  3 \\
\hline
\end{tabular}
}
\noindent
\end{center}
{\scriptsize Note.-Col. (1): Object name. Cols. (2)-(8): Emission line ratios 
relative to H$\beta$ flux. \\
References.-(1) Morganiti et al. 1997, (2) Gelderman \& Whittle 1994, 
(3) Lawrence et al. 1996, (4) Labiano et al. 2005
}
\end{table}

\clearpage
\begin{figure}
\plotone{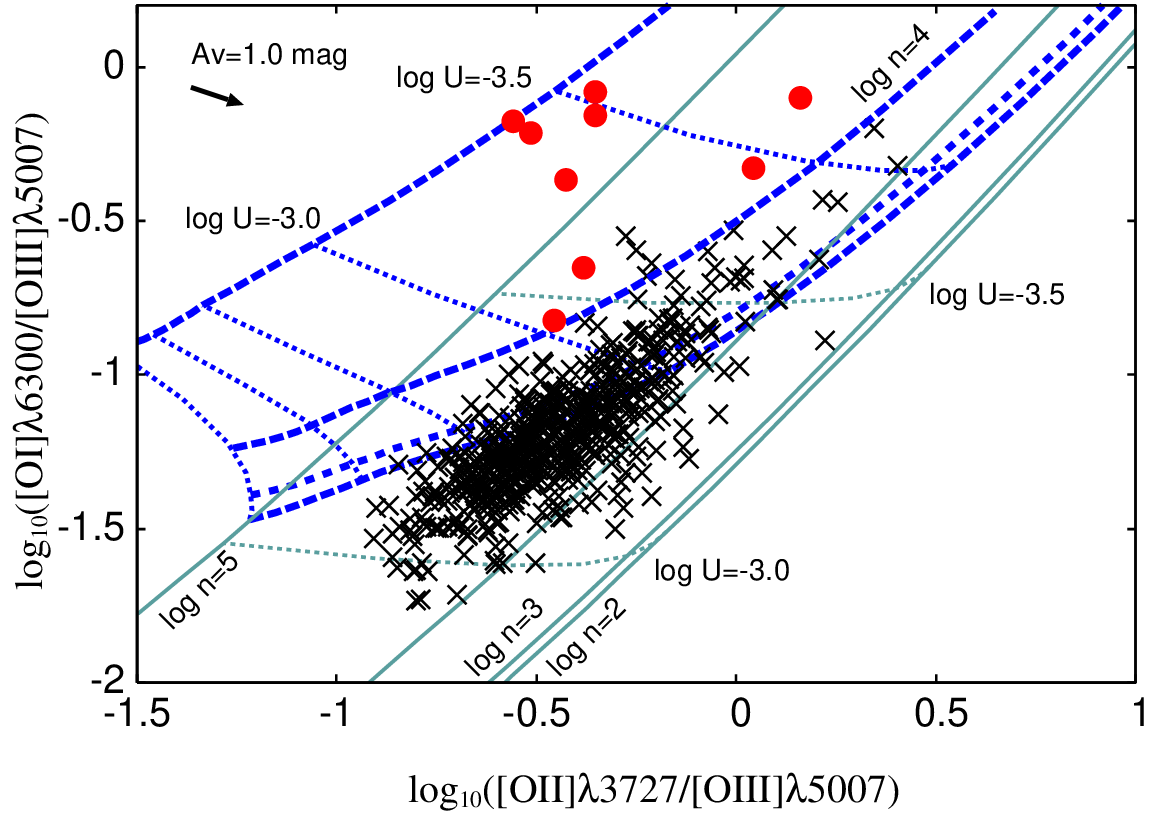}
\caption{
Diagram of [O{\sc i}]$\lambda$6300/[O{\sc iii}]$\lambda$5007 vs. 
[O{\sc ii}]$\lambda$3727/[O{\sc iii}]$\lambda$5007. 
The observed line ratios from the literature are presented for
powerful YRL AGNs (red circles) and radio-quiet Seyfert 2 galaxies (black crosses).
Solid and dashed lines represent our photoionization model calculations using
the SED with BBB and the SED without BBB, respectively. 
}
\end{figure}

\begin{figure}
\plotone{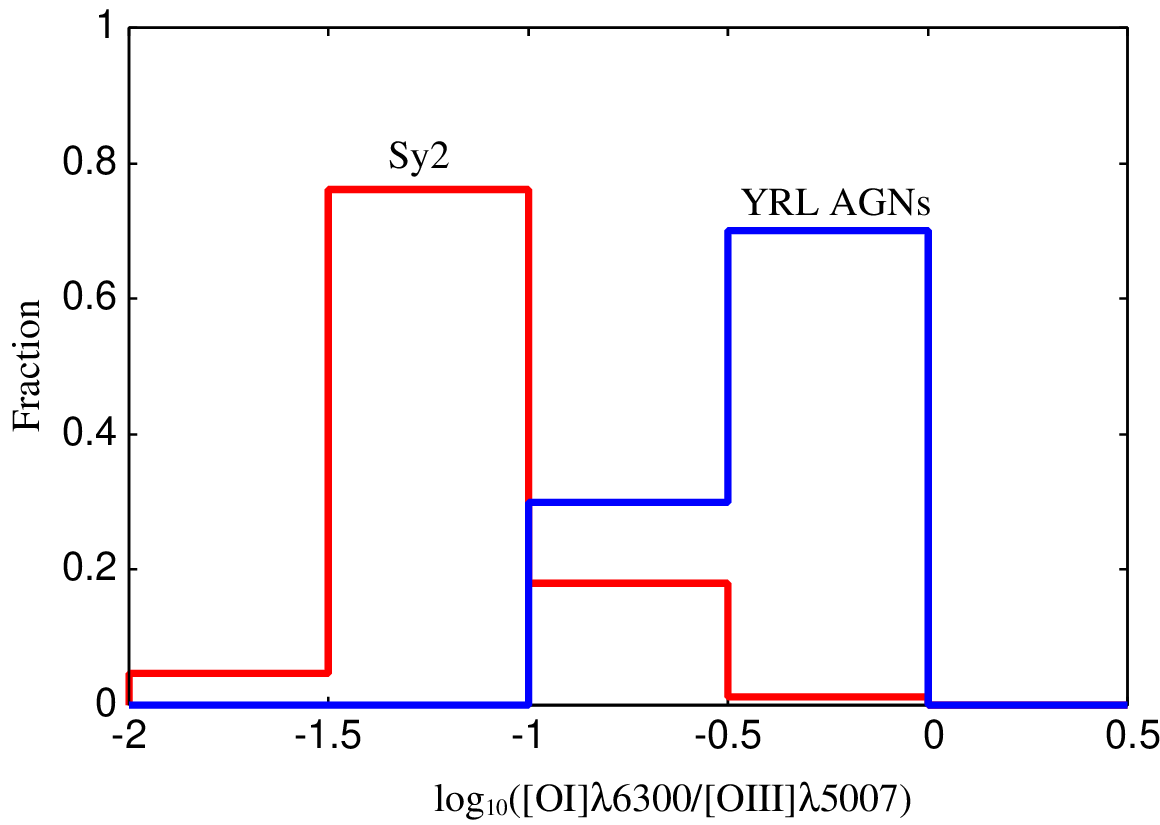}
\caption{
Frequency distributions of emission line flux ratios 
[O{\sc i}]$\lambda$6300/[O{\sc iii}]$\lambda$5007, for powerful YRL AGNs (red) 
and Seyfert 2 galaxies (blue).
}
\end{figure}

\begin{figure}
\plotone{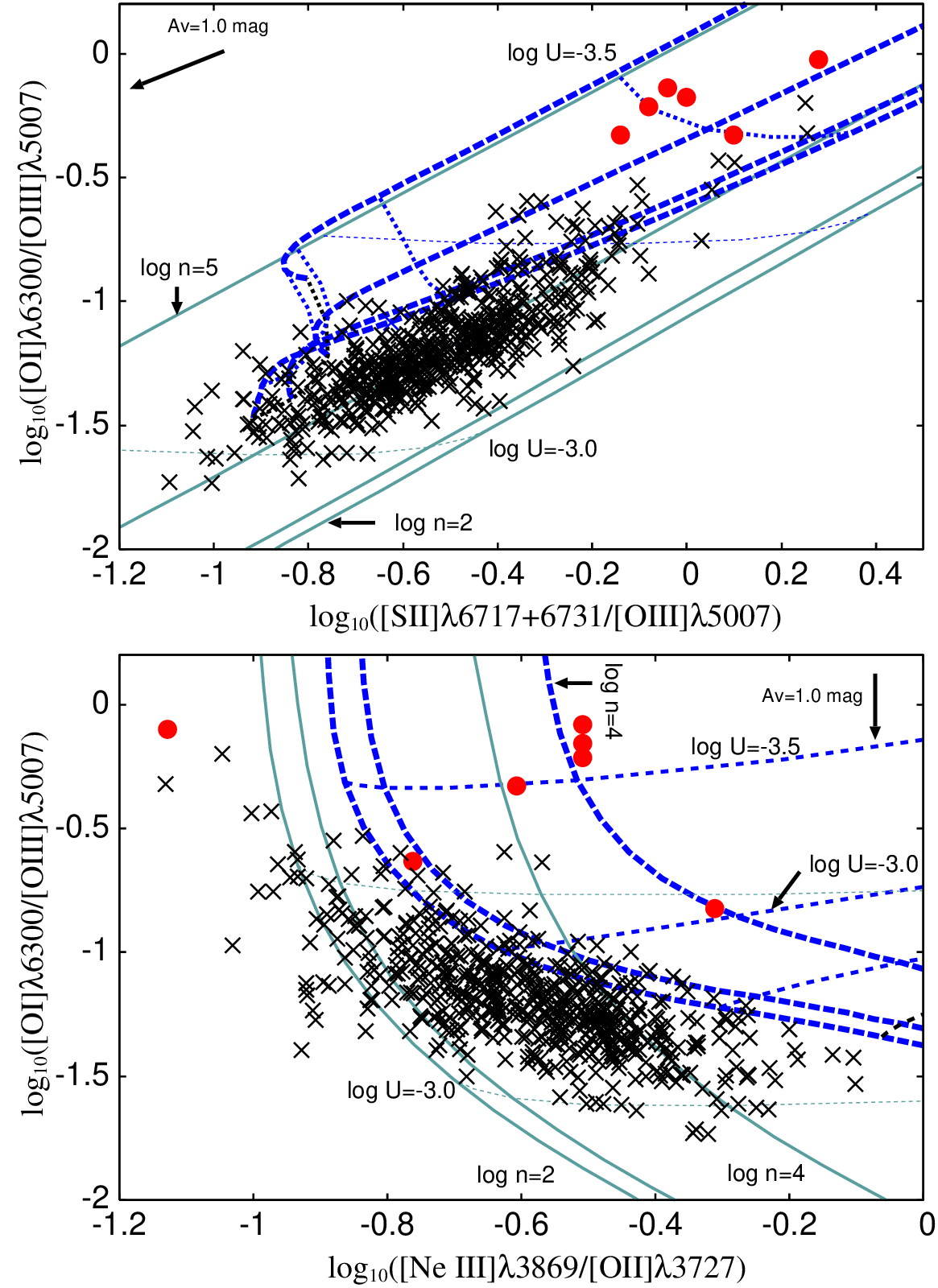}
\caption{\small 
Diagram of [O{\sc i}]$\lambda$6300/[O{\sc iii}]$\lambda$5007 vs. 
[S{\sc ii}]$\lambda$6717+6731/[O{\sc iii}]$\lambda$5007 (top) 
and that of [O{\sc i}]$\lambda$6300/[O{\sc iii}]$\lambda$5007 vs. 
[Ne {\sc iii}]$\lambda$3869/[O{\sc iii}]$\lambda$3727 (bottom). 
Symbols and lines are the same as Figure 1.
}
\end{figure}

\begin{figure}
\begin{center}
\plotone{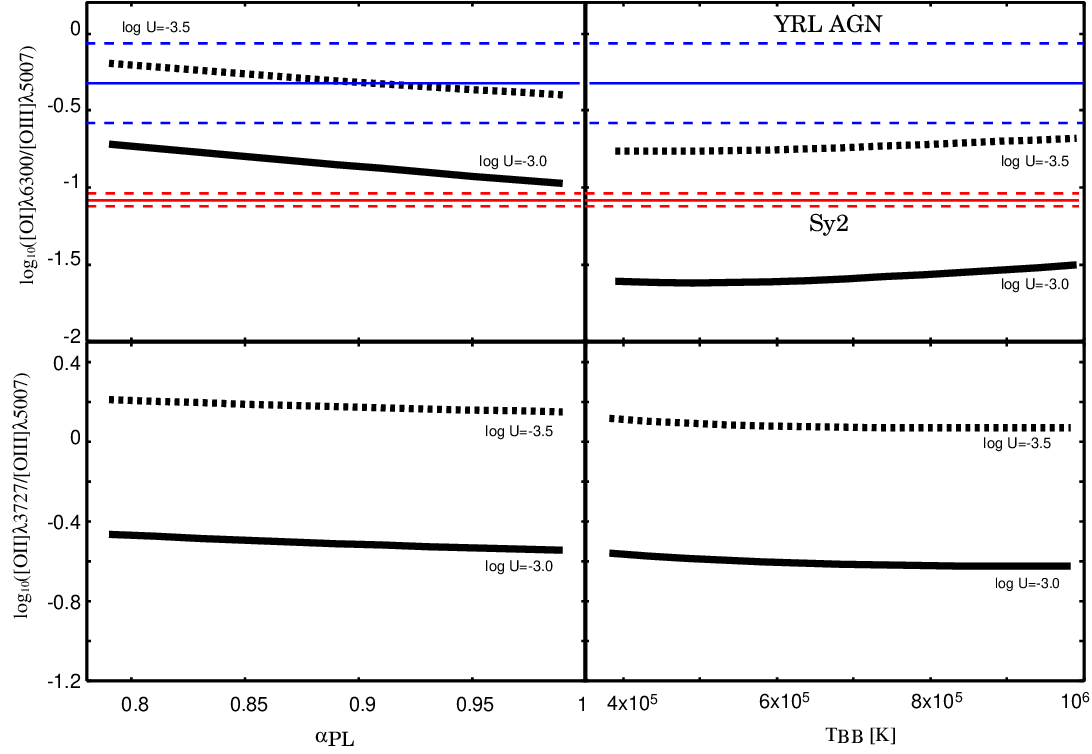}
\end{center}
\caption{
The optical emission line ratios ([O{\sc i}]$\lambda 6300$/[O{\sc iii}]
$\lambda 5007$ and [O{\sc ii}]$\lambda 3727$/[O{\sc iii}]$\lambda 5007$) 
as a function of the power-law index of RIAF-like SEDs, $\alpha_{\rm PL}$ (left), 
and the characteristic temperature, $T_{\rm BB}$ (right) 
for a standard accretion disk, 
assuming $n_{\rm H}=10^{4}\,{\rm cm}^{-3}$. 
The solid lines represent the case of $\log U=-3.0$, while 
the dotted lines denote that of $\log U=-3.5$. 
The blue (red) solid line represents the mean value of 
$\log($[O{\sc i}]$\lambda$6300/[O{\sc iii}]$\lambda$5007) 
for powerful YRL AGNs (Seyfert 2 galaxies). 
The blue (red) dotted lines correspond to the dispersion of 
$\log($[O{\sc i}]$\lambda$6300/[O{\sc iii}]$\lambda$5007).
}
\end{figure}

\begin{figure}
\plotone{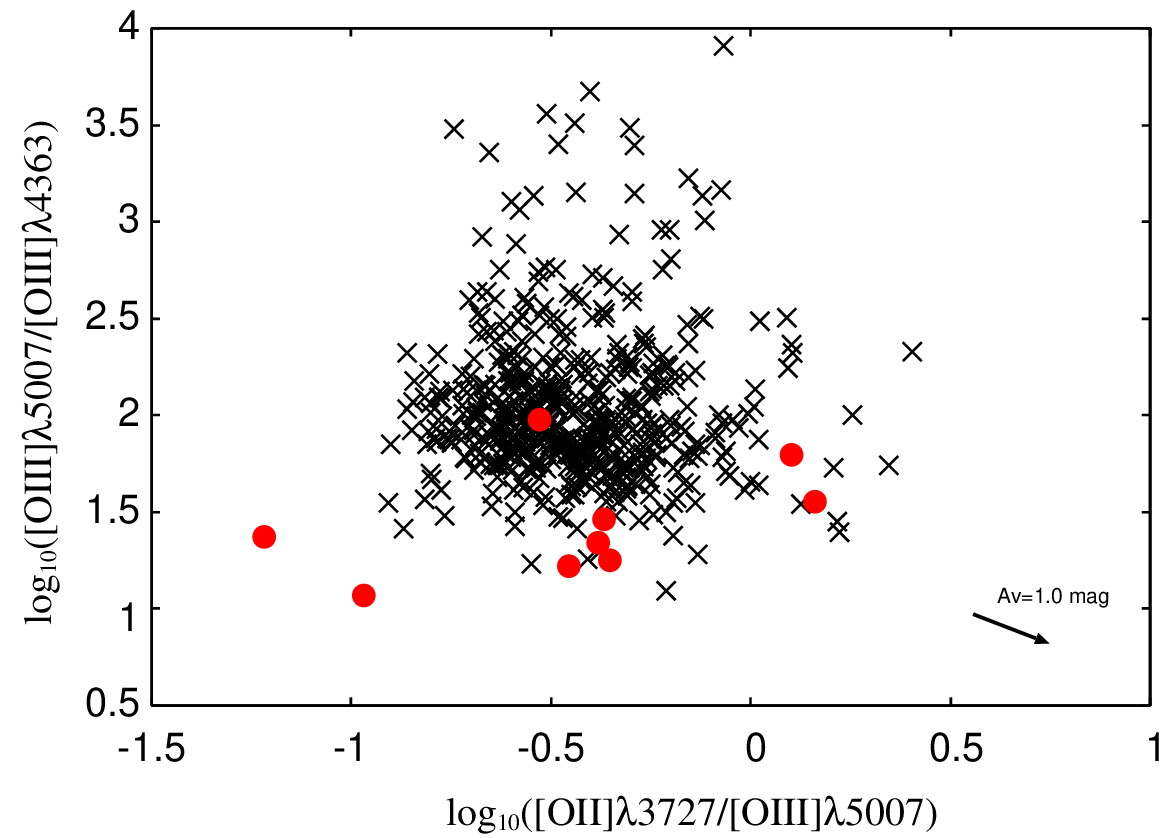}
\caption{
Diagram of [O{\sc iii}]$\lambda$5007/[O{\sc iii}]$\lambda$4363 vs. 
[O{\sc ii}]$\lambda$3727/[O{\sc iii}]$\lambda$5007. 
Symbols and lines are the same as in Figure 1.
}
\end{figure}

\begin{figure}
\plotone{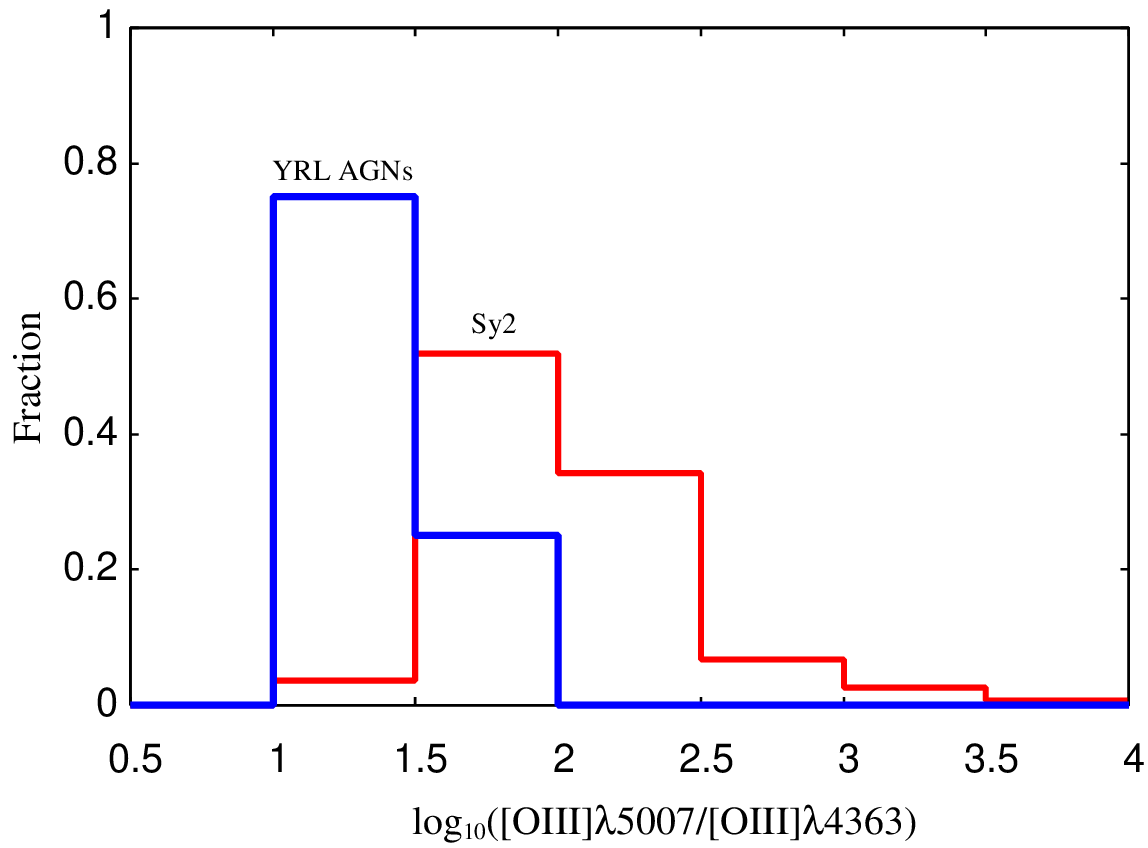}
\caption{
Frequency distributions of the emission line flux ratios of 
[O{\sc iii}]$\lambda$5007/[O{\sc iii}]$\lambda$4363, for powerful 
YRL AGNs (red) and Seyfert 2 galaxies (blue).
}
\end{figure}

\begin{figure}
\begin{center}
\includegraphics[width=8cm]{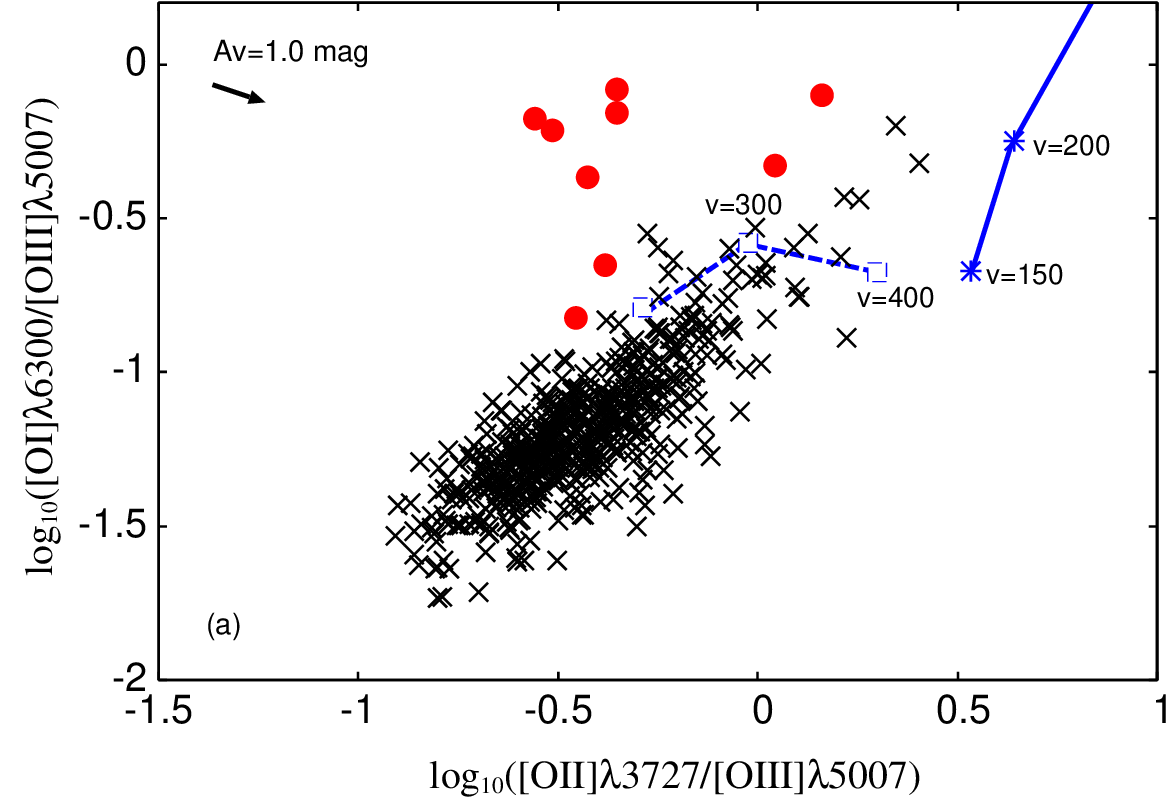}
\includegraphics[width=8cm]{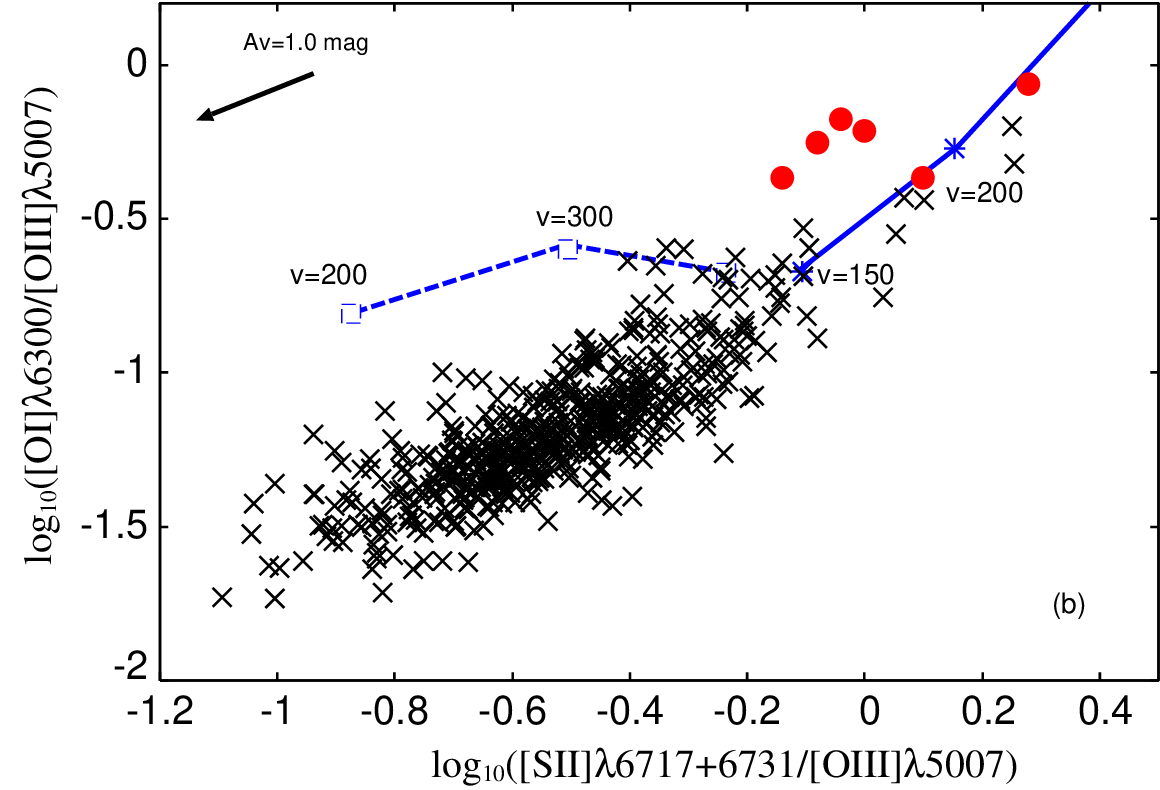} \\
\includegraphics[width=8cm]{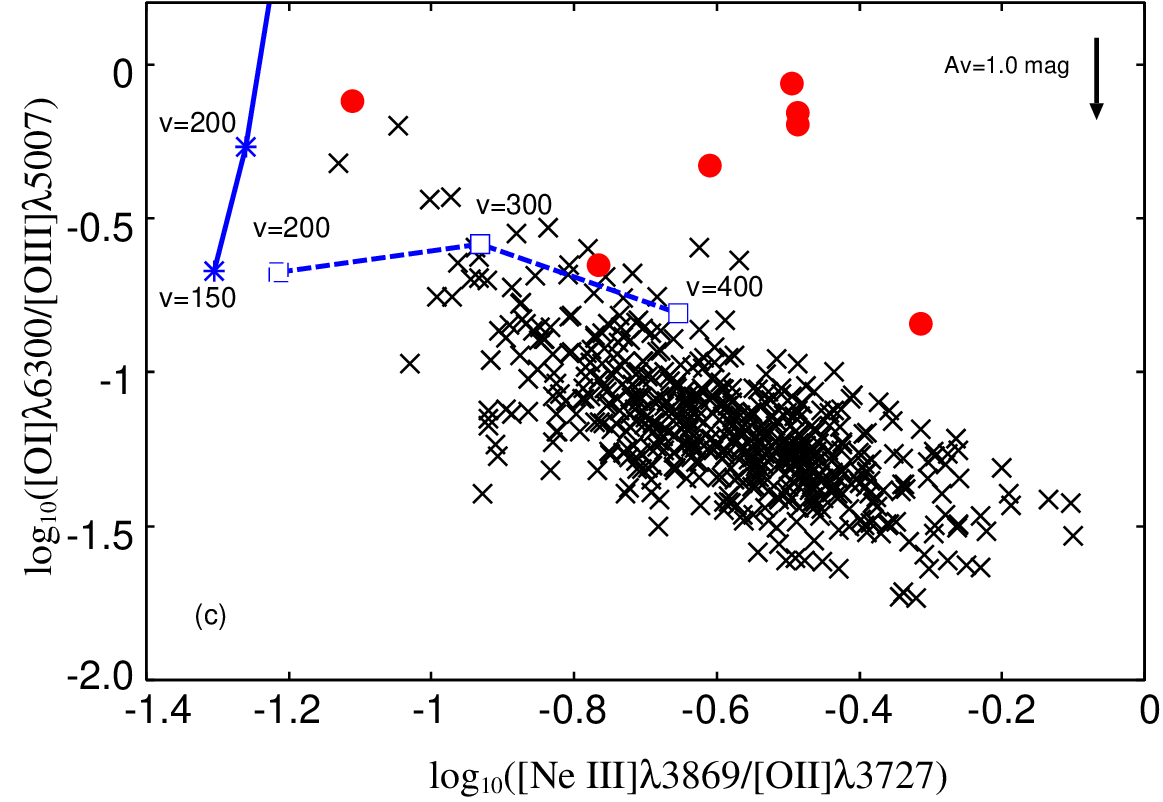}
\end{center}
\caption{
(a) Diagram of [O{\sc i}]$\lambda$6300/[O{\sc iii}]$\lambda$5007 vs. 
[O{\sc ii}]$\lambda$3727/[O{\sc iii}]$\lambda$5007, 
(b) diagram of [O{\sc i}]$\lambda$6300/[O{\sc iii}]$\lambda$5007 vs. 
[S{\sc ii}]$\lambda$6717+6731/[O{\sc iii}]$\lambda$5007, 
and (c) diagram of [O{\sc i}]$\lambda$6300/[O{\sc iii}]$\lambda$5007 vs. 
[Ne{\sc iii}]$\lambda$3869/[O{\sc iii}]$\lambda$3727.
Predictions of the shock model (Dopita \& Sutherland 1995) 
are represented by blue lines. 
The blue solid lines denote the pure-shock model 
($150\, {\rm km/s} \leq v_{\rm shock} 
\leq 500\, {\rm km/s}$), while the blue dotted lines denote the models 
considering the effect of precursor ($200\, {\rm km/s} \leq v_{\rm shock} 
\leq 500\, {\rm km/s}$). Chemical composition of solar abundances and the 
magnetic parameter of $B/\sqrt{n}=2\mu G\,{\rm cm}^{2/3}$ are assumed 
in these models (see Table 1 in Dopita \& Sutherland 1995 for details). 
Symbols and line are the same as Figure 1.
}
\end{figure}

\begin{figure}
\plotone{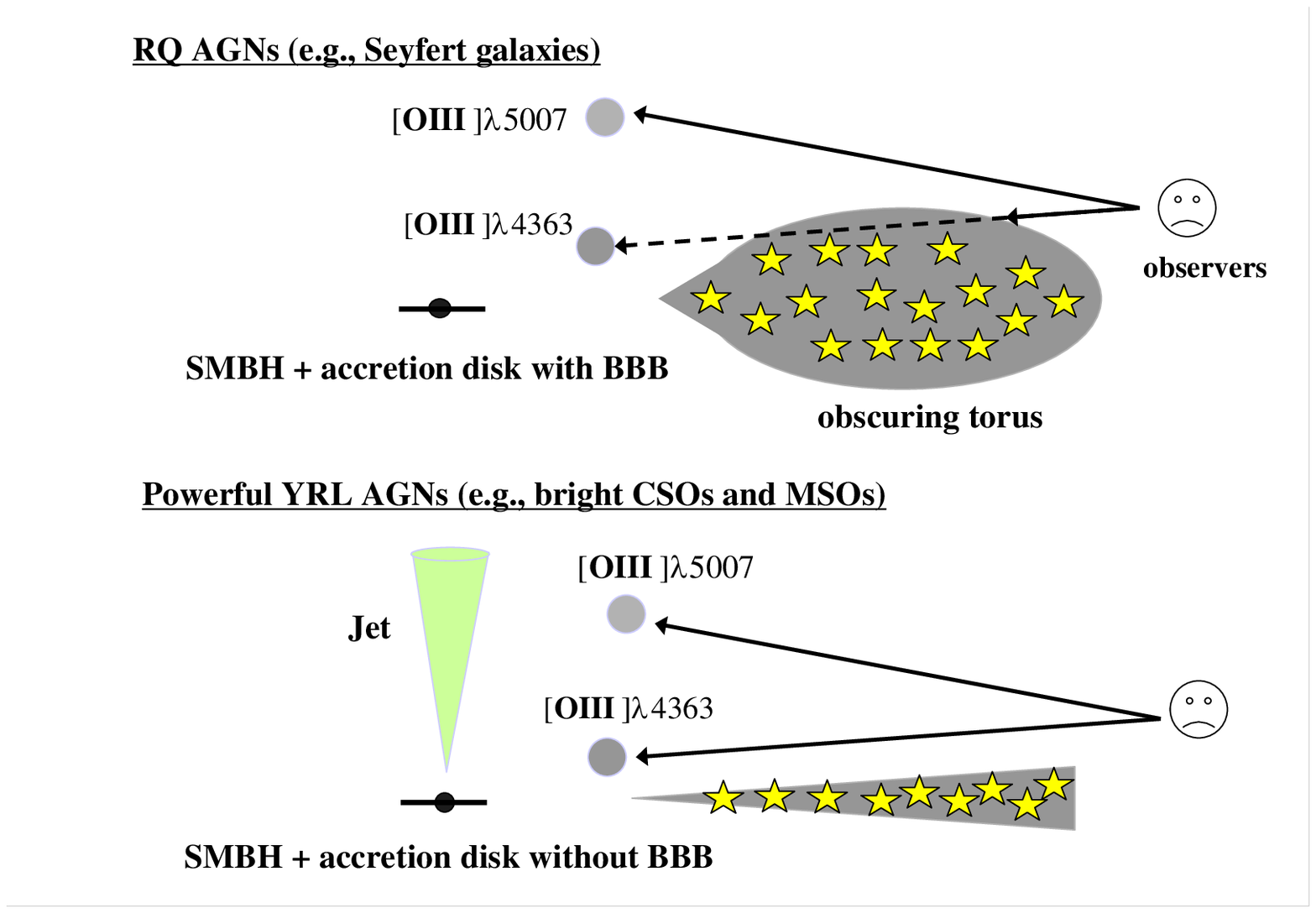}
\caption{
Schematic sketches for the properties of accretion disk and the 
obscuring materials of the RQ Seyfert 2 galaxies (upper) 
and the powerful YRL AGNs, i.e., CSOs and MSOs (lower). 
The light gray circles denote the narrow line clouds emitting 
[O{\sc iii}]$\lambda$5007 and the dark gray circles represent the narrow line 
clouds emitting [O{\sc iii}]$\lambda$4363.
}
\end{figure}

\end{document}